\documentstyle[psfig,prd,preprint,aps]{revtex}
\tightenlines
\newcommand{\be}{\begin{equation}}
\newcommand{\ee}{\end{equation}}
\newcommand{\ba}{\begin{eqnarray}}
\newcommand{\ea}{\end{eqnarray}}
\begin{document}
\title{\bf Non-BPS Dyons and Branes in the Dirac-Born-Infeld Theory}

\author{H.R.~Christiansen,\\
{\normalsize\it Centro Brasileiro de Pesquisas F\'\i sicas, CBPF - DCP}\\
{\normalsize\it Rua Dr. Xavier Sigaud 150, 22290-180 Rio de Janeiro, Brazil}\\
~
\\
N.~Grandi
, 
F.A.~Schaposnik\thanks{Associated with CICBA} \, and 
G.~Silva 
\\
{\normalsize\it
Departamento de F\'\i sica, Universidad Nacional de La Plata}\\
{\normalsize\it
C.C. 67, 1900 La Plata, Argentina}}

\maketitle
 
 \vspace{- 7 cm}
\hspace{12.2 cm}  La Plata-Th 99/13
\vspace{6.9 cm}

\begin{abstract}
Non-BPS dyon solutions to D3-brane actions are constructed when
one or more scalar fields describing transverse fluctuations
of the brane, are considered. The picture emerging from such non-BPS
configurations is analysed, in particular the response of the 
D-brane-string system to small perturbations.
\end{abstract}
\date{}

%


 
\section{Introduction}
Solutions to Dirac-Born-Infeld (DBI) theory have recently drawn much
attention in connection to the dynamics of Dp-branes \cite{Pol}-\cite{Ts}.
Indeed,
the DBI action for $p+1$ dimensional gauge fields and a number of scalars 
describing transverse fluctuations of the brane allow static 
Bogomol'nyi-Prasad-Sommerfield (BPS) and non-BPS configurations, 
which can be interpreted in terms of branes and strings attached to them.

Although many (static) properties related to intersecting branes come from
supersymmetry and BPS arguments, specific dynamical features depend
strongly on the non-linearity of the DBI action. In particular, those
related to the effective boundary conditions imposed to strings attached 
to branes must be investigated using the full DBI action. Moreover, 
non-BPS configurations might be useful for the study of certain
 non-perturbative
aspects of field theories that describe the low energy dynamics 
of branes\cite{BG}.

BPS   and non-BPS (throat) purely electric solutions 
to DBI theory were constructed in \cite{CM}-\cite{G}.
Also the 
propagation of a perturbation normal to both the string and  the 3-brane
action was investigated in \cite{CM}
for a BPS background. The results obtained  
show that the picture of a string attached
to the brane with Dirichlet boundary conditions emerges
naturally from DBI dynamics. 
In \cite{Svv}-\cite{Svv2} 
perturbations polarized  along  the brane  in a BPS background
were studied and it was shown  that Neumann boundary 
conditions are realized in this case.

Other purely electric non-BPS solutions 
to DBI action for the world volume gauge field 
and scalar fields  were constructed in \cite{Hashi} where also magnetically
charged BPS solutions were discussed. A detailed study of  BPS dyonic
solutions  was   presented later in \cite{BLM}.
 
In this paper we concentrate in the case of D3-branes and explicitly 
construct non-BPS dyon solutions when the U(1) gauge field couples to one
or more scalar fields. We then analyse the solutions 
in connection with the geometry of the bending of the
brane due to the tension  of a $(n,m)$-string \cite{Sch}-\cite{Wi} 
carrying both electric 
and magnetic charges . Studying the energy of these non-BPS 
configurations, we
compare the results with those obtained in the purely electric
BPS and non-BPS cases \cite{CM}-\cite{BLM}. We also study small
excitations, transverse both to the brane and to the string, in order to
test whether the response of the non-BPS solution is consistent with the
interpretation  in which the brane-string system described  
corresponds to the appropriate (Dirichlet) boundary condition.

The plan of the paper is as follows: in section II we construct the
non-BPS  solutions, with both electric and magnetic
charges, to the  DBI model for an Abelian gauge field in the 
world volume, coupled to one scalar. We discuss the properties of the 
solutions and compare  them
to other solutions
 already described in the literature. We also compute the 
renormalized energy and interpret the dyonic non-BPS
solution in terms of  strings attached to  D3-branes. Then, in section
III, we consider small perturbations to the non-BPS
background, normal to the brane and to the string,
in order to test the resulting  boundary conditions. Finally, we
summarize and discuss our results  in section IV.

\section{Solutions of the Dirac-Born-Infeld Action}
 The D3-brane action in the static gauge takes the form  
\be
S = -T_3 \int d^4x \sqrt{-{\rm det} (\eta_{\mu \nu} + T^{-1} F_{\mu \nu} + 
\partial_\mu X^a \partial_\nu X^a)}
\label{D3}
\ee
where $\eta_{\mu \nu}$ is Minkowski metric in $3+1$ dimensions, 
${\rm diag}(\eta_{\mu \nu}) = (-1,1,1,1)$,
$F_{\mu \nu}$ is the world volume electromagnetic  gauge field strength,
$X^a$ are scalar fields ($a = 4,5,\ldots,9$)  which describe 
transverse fluctuations of the
brane
and 
\be
T_3 = \frac{1}{2\pi g_s} T^2  \, , \;\;\; T = \frac{1}{2\pi \alpha'}
\label{T3}
\ee
with $g_s$ the string coupling constant.
This action can be obtained by dimensional reduction  of a Born-Infeld action
in 10 dimensional
flat space-time ($x^M$, $M=0,1,2, \ldots, 9$), assuming that the fields 
depend only 
on the first $1+3$ coordinates $x^\mu$ and that the extra components 
$A_4, A_5 \ldots
A_9$ of the gauge
field represent the scalar fields. 
We shall  consider first the case in which there is just one excited
scalar field,
$X^a = \delta^{a9} X$.  In this case
eq.(\ref{D3}) takes the form
\begin{eqnarray}
S &=& -T_3 \int d^4x  \left( \left(1 + \partial_\mu X\partial^\mu X \right)
 \left(1
+ \frac{1}{2T^2} F_{\mu \nu} F^{\mu \nu}\right)
 - \frac{1}{16 T^4} 
\left(\tilde F_{\mu \nu} F^{\mu \nu}
\right)^2   \right.  \nonumber\\
&&  
\left.  +
\frac{1}{T^2}\partial^\mu XF_{\mu \nu} F^{\nu \rho}\partial_\rho X
\right)^{1/2}
\end{eqnarray}

The equations of motion for time-independent solutions read,
\begin{eqnarray}
&&\vec \nabla \cdot \left( \frac{1}{R} \left( \vec \nabla X + \frac{1}{T^2}
(\vec B \cdot  \vec \nabla X) \vec B +\frac{1}{T^2} \vec E \wedge (\vec E \wedge 
\vec \nabla X)
\right)\right) = 0\nonumber\\
&&\vec \nabla \cdot \left( \frac{1}{R} \left( \vec E 
+\vec \nabla X \wedge (\vec E \wedge 
\vec \nabla X)
 + \frac{1}{T^2} (\vec E \cdot \vec B) \vec B
\right)\right) = 0\nonumber\\
&&\vec \nabla \wedge \left( \frac{1}{R} \left( \vec B 
+  (\vec B \cdot 
\vec \nabla X) \vec \nabla X
 -\frac{1}{T^2} (\vec E \cdot \vec B) \vec E
\right)\right)= 0\label{tch} 
\end{eqnarray}
Here, the $U(1)$ electric field $\vec E$ and magnetic induction
$\vec B$  are defined as usual as
\be
 E_i = F_{i0} \; , \;\;\;  B_i = \frac{1}{2} \varepsilon_{ijk}F_{jk} 
\label{EB}
\ee
Concerning $R$, it is defined as
\begin{eqnarray}
R^2 &=& 
           1
         +
           (\vec \nabla X)^2
         +
           \frac{1}{T^2}
           \left(
           	    \vec B^2 
                  - 
                    \vec E^2
                  +
                    (\vec B \cdot \vec \nabla X)^2
                  -
                    (\vec E \wedge \vec \nabla X)^2
            \right)
           \nonumber\\
           && -\frac{1}{T^4}
            (\vec E \cdot \vec B)^2  
\label{R}
\end{eqnarray}

Now, since we are interested in bion solutions \cite{CM}-\cite{G} carrying 
both electric $q_e$  and magnetic $q_m$
charges, this necessarily implies that $\vec E$ and $\vec B$ 
have delta function sources (c.f. \cite{GK} where
dyon configurations with an extended magnetic
source are constructed). For the case $X = 0$ one easily finds  
such a solution to (\ref{tch}) in the form
\be
A_0 = -\frac{q_e}{4\pi} \int_r^\infty dr
 \frac{1}{\sqrt{(q_e^2 + q_m^2)(4\pi T)^{-2} + r^4}}
\label{ab}
\ee
\be
A_\varphi = \frac{q_m}{4\pi r} \frac{(1 - 
\cos \theta)}{\sin \theta} 
\; , \;\;\; A_\theta = A_r = 0
\label{ba}
\ee
This solution can be obtained from the purely electric Born-Infeld 
one by a duality rotation \cite{GR}.

Following \cite{G}-\cite{Hashi} , we   construct the general solution
by performing a
boost (in $10$ dimensional space)  in the $x^9$ direction leading to 
\begin{eqnarray}
X &=& -\frac{q_e\sqrt a}{4\pi T} \int_r^\infty dr \frac{1}{\sqrt{r^4_0 + r^4}}
\nonumber\\
A_0 &=&  -\frac{q_e}{4\pi} \int_r^\infty dr \frac{1}{\sqrt{r_0^4+ r^4}}
\nonumber\\
A_\varphi &=& \frac{q_m}{4\pi r} \frac{(1 - \cos \theta)}{\sin \theta} 
\; , \;\;\; A_\theta = A_r = 0
\label{xab}
\end{eqnarray}
where
\be
r_0^4 = (4\pi T)^{-2}((1-a)q_e^2 + q_m^2)
\label{cerraria}
\ee
and  $a$ is related to the square of the boost velocity. 
The electric and magnetic fields associated with (\ref{xab}) take the
form
\begin{eqnarray}
\vec E &=& \frac{q_e}{4\pi\sqrt{ r_0^4 + r^4}} \check r \nonumber\\
\vec B &=&  \frac{q_m}{4\pi r^2}  \check r
\label{BE}
\end{eqnarray} 
Note that since the boost is in the $x^9$ direction, it does not affect
the transverse $x^i \, , i=1,2, \ldots,8$ directions. Moreover, since
we are considering static solutions, $A_\varphi$ in (\ref{xab}) is not affected
and the magnetic field remains unchanged by the boost.

These solutions generalize all known (one source) solutions 
already discussed in the
DBI-brane context, either the BPS or non-BPS ones.  
Setting $q_m=0$ we recover, for $a<1$,
the electric bions, for $a>1$, the electric
throat/catenoid solutions
 and, for $a=1$, the electric BPS
solution \cite{CM}-\cite{Hashi}. 
The new solutions we have found generalize these electric bions and
throats to  dyonic ones. 

Concerning the  magnetic field $\vec B$,
 it is important
to note that its value is $a$ independent and has the usual Dirac monopole
functional form corresponding to a quantized
magnetic charge. Indeed, the magnetic induction $\vec H$  is given by
\be
\vec H \equiv \frac{1}{R} 
\left( \vec B + \left(
\vec B \cdot \vec \nabla X
\right) \vec \nabla X - \frac{1}{T^2}\left(\vec E \cdot \vec B
\right) \vec E
\right) = \frac{q_m}{4\pi\sqrt{ r_0^4 + r^4}} \check r  
\label{HH}
\ee
The electric   and magnetic  
charges of the solutions  were  
adjusted so that
\be
 \int_{S_\infty} dS_i E^i = q_e  \, , \;\;\;\;   \;\;\;\;  \int
 _{S_\infty} dS_i B^i = q_m
\label{cargas}
\ee 
It is useful to define the scalar field charge $q_s$ in the form
\be
q_s \equiv T \int_{S_\infty} dS_i \partial^i X = \sqrt a \, q_e
\label{lades}
\ee
In terms of  charges, $r_0$ in (\ref{cerraria}) takes the form
\be
r_0^4 = (4\pi T)^{-2} \left(q_e^2 +q_m^2 - q_s^2\right)
\label{qq}
\ee
 
From (\ref{xab}) one can see that,
in the range $q_s^2 \leq q_m^2 + q_e^2$, the scalar  takes essentially
the form depicted in Fig. 1. Qualitatively, its behavior is similar
 to
the purely electric solution found in \cite{Hashi} except that the
existence of a non-zero magnetic charge lowers the height of the cusp.

For $q_s^2> q_e^2 +q_m^2$ the scalar takes the form of the  solution
depicted in Fig. 2 which can be viewed as two asymptotically flat branes
(in fact, a brane-antibrane pair)
joined by a throat of radius $r_t$. These branes
are separated a distance $\Delta = 2 |X(r_t)|$ which corresponds to
the difference between the  two asymptotic values of $X(r)$.  
The radius of the throat is given by
\be
r_t^4 = -r_0^4 = (4\pi T)^{-2}\left(q_s^2 - q_e^2 - q_m^2\right)
\label{th}
\ee
Note that increasing the magnetic charge makes the throat slimmer and 
$\Delta$ larger.

BPS solutions correspond to the case $r_0 = 0$. That is,
when the scalar charge satisfies
\be
q_s^{BPS} = \pm \sqrt{q_e^2 + q_m^2}
\label{has}
\ee
 When (\ref{has}) holds, 
the
solutions satisfying  Bogomol'nyi equations   
\begin{eqnarray}
\vec E &=& T \cos \xi \vec \nabla X\label{quese}
\\
\vec B &=& T\sin \xi \vec \nabla X
\label{quesecalle}
\end{eqnarray}
are
\begin{eqnarray}
X&=& - \frac{\sqrt{q_e^2 + q_m^2}} {4\pi T r} \nonumber\\
\vec E &=& \frac{q_e}{4\pi r^2}\check r\nonumber\\ 
\vec B &=& \frac{q_m}{4\pi r^2}\check r
\label{grandiestadormido}
\end{eqnarray}
with
\begin{eqnarray}
\cos \xi = \frac{q_e}{\sqrt{q_e^2 + q_m^2}}    
\label{cadacosa}
\end{eqnarray}
Note that $r_0 = 0$ implies that the boost parameter $a = 1 + q_m^2/q_e^2 >1$. 
In particular,
for $q_m=0$ the boost is a light-cone one. In fact, $q_m = 0$  corresponds
to $\xi = 0$
and then
(\ref{xab}) reduces  
to the electric BPS solution discussed in  \cite{CM}-\cite{G}. 
The
choice $\xi = \pi/2$ ($q_e=0$) corresponds to the magnetic BPS solution discussed in 
\cite{G}. For arbitrary $\xi$ our BPS solution coincides with that
analysed in \cite{BLM}. 

Let us now compute the static energy for the non-BPS
configurations described earlier
and relate it to the bending of branes. We consider the case 
in which there is only one D3-brane and first compute the energy stored in
the world volume of the brane for the configuration 
(\ref{xab})-(\ref{BE}) with $r_0^4 \geq 0$,
\begin{eqnarray}
E_{wv} &=& \int d^3x \, T_{00} \nonumber\\
&=& \frac{2}{g_s} \int r^2 dr \left(
  \frac{q_e^2 + q_m^2 }{(4\pi r)^2\sqrt{r^4 + r_0^4}} + T^2 \left(
 \frac{r^2}{\sqrt{r^4 + r_0^4}} - 1\right) \right)\nonumber\\
 &=& \frac{\Gamma^2(1/4)}{96 \pi^{5/2} g_s}
 \left(  {2}  (q_e^2 + q_m^2) + q_s^2 \right) \frac{1}{r_0}
 \label{divergetodo}
 \end{eqnarray}
Since the BPS limit  is reached when $r_0 = 0$,  we
see  that $E_{wv}$ diverges precisely at the point which
should correspond to the lower
bound for the energy. In order to avoid this problem, one
can normalize the energy with respect to the Bogomol'nyi value. To
this end we define
\begin{eqnarray}
E &=&  E_{wv} - E_{sub} = E_{wv} - \frac{\Gamma^2(1/4)}{ 32 \pi^{5/2} g_s }
|q_s| \sqrt{q_e^2 + q_m^2} \frac{1}{r_0} \nonumber\\
&=& 
 \frac{\Gamma^2(1/4)}{6 \sqrt \pi  g_s}T^2
 \left( 2 - \frac{3|q_s|}{|q_s| + \sqrt{q_e^2 + q_m^2}   }
 \right) r_0^3
 \label{sub}
\end{eqnarray}

Clearly,  $E=0$ in the BPS case ($r_0 = 0$). In general, 
$0 \leq E < \infty$ and then the BPS configuration 
gives the lower bound for the energy. We show in Fig. 3 the
energy as given by (\ref{sub}) as a function of the scalar charge. At
fixed electric charge, one can see that, as the magnetic charge
grows,  the Bogomol'nyi bound is attained for larger scalar charge.
 
The subtraction performed in (\ref{sub}) can be interpreted as follows.
Using eq.(\ref{xab}),  $E_{wv}$, as given by eq.(\ref{divergetodo}),
can be written as
\be
E_{wv} =    \frac{T}{6\pi g_s}  \frac{1}{|q_s|} \left(2(q_e^2 + q_m^2) + q_s^2 \right)
\, |X(0)|
\label{lar}
\ee
Concerning the subtracted term, it takes the form
\be
E_{sub}
=  \frac{T}{2\pi g_s} \sqrt{q_e^2 + q_m^2} \, |X(0)|
\label{latex}
\ee
The connection between   the dyon electric field and fundamental strings
leads to the quantization of the electric flux \cite{CM} so that  
$q_e =2\pi g_s n $.
For the magnetic charge, we write $q_m = 2\pi m$. Then, $E_{sub}$ can be
rewritten in the form
\be
E_{sub}
=  T   \sqrt{n^2 + \frac{1}{g_s^2}m^2} |X(0)|
\label{nm}
\ee
The renormalized energy 
$E$ as defined in (\ref{sub}) can then be written as
\begin{eqnarray}
E &=& 
E_{wv} - T_{(n,m)}\int_0^{|X(0)|} dX \label{joder}\\
&=& E_{wv} + T_{(n,m)} \int_{|X(0)|}^\infty dX  - 
T_{(n,m)} \int_0^\infty dX 
\label{tbdral}
\end{eqnarray}
where
\be
T_{(n,m)} =  T \sqrt{n^2 + \frac{1}{g_s^2}m^2}
\label{sss}
\ee
Formula (\ref{tbdral}) makes clear the rationale of the subtraction: 
the second  term in the r.h.s. of (\ref{tbdral}) represents the energy
of a semi-infinite string (with tension $T_{(n,m)}$) extending from the cusp
of the spike to infinity. The third term subtracts the infinite energy
of a string extending from $0$ (from a flat brane)  to infinity. We
are then computing the energy of a brane pulled by a string with respect
to the energy of a non-interacting configuration brane+string
(which turns to be a BPS solution).

We represent in figure 4 a sequence of growing spikes as the scalar
charge increases up to the point it attains its BPS value 
$q_s^{BPS}$.

One can also compute the static energy stored in the worldvolume
for the throat
solution ($r_t^2 =-r_0^2 >0$).  One gets
\be
E_{wv} = \frac{\Gamma^2(1/4)}{48\sqrt 2 \pi^{5/2}g_s}
\left(  {2}  (q_e^2 + q_m^2) + q_s^2 \right) \frac{1}{r_t} + 
\frac{4}{3}\frac{T^2}{g_s} r_t^3
\label{eth}
\ee
 which also diverges in the BPS ($r_t \to 0$) limit.
The adequate subtraction now required in order to get a finite
result  is
\be
E_{sub}= \frac{\Gamma^2(1/4)}{ 16\sqrt 2 \pi^{5/2} g_s }
|q_s| \sqrt{q_e^2 + q_m^2} \frac{1}{r_t}
\label{susm}
\ee
We then have for the throat
 \be
 E =  E_{wv} - E_{sub} = \frac{4}{3} \frac{T^2}{g_s} 
 \left(1 + 
 \frac{\Gamma^2(1/4)}{4 \sqrt{2\pi}}
 \frac{|q_s| -
  2\sqrt{q_e^2 + q_m^2}}
  {|q_s| +\sqrt{q_e^2 + q_m^2}}
 \right)r_t^3
 \label{beg}
 \ee
 The subtracted energy $E_{sub}$ defined by eq.(\ref{susm}) can be written as
 \be
 E_{sub} =  \frac{T}{2\pi g_s}\sqrt{q_e^2 + q_m^2} 2\vert X(r_t) | = 
 T_{(n,m)} \Delta
 \label{bis}
 \ee
where $T_{(n,m)}$ is defined by eq.(\ref{sss}). Then, 
the finite energy $E$ in (\ref{beg}) corresponds to the difference between
the throat solution and a non-interacting configuration brane-string-antibrane.
 
We shall now briefly describe non-BPS solutions of the DBI action 
when two
scalar fields are present. Starting from action (\ref{D3}) with
$X^8 = X$ and $X^9 = Y$, the corresponding equations of motion read
\begin{eqnarray}
&&\vec \nabla \cdot \left( \frac{1}{R} \left( \vec \nabla X 
+
(\vec \nabla X \wedge \vec \nabla Y)\wedge \vec \nabla Y
-
\frac{1}{T^2}
\vec E \cdot (\vec \nabla X \wedge \vec \nabla Y) (\vec E \wedge \vec \nabla Y)
\right.\right.
\nonumber\\&& \left.\left.
+ \frac{1}{T^2}
(\vec B \cdot  \vec \nabla X) \vec B +\frac{1}{T^2} \vec E \wedge (\vec E \wedge 
\vec \nabla X)
\right)\right) = 0\nonumber\\
&&\vec \nabla \cdot \left( \frac{1}{R} \left( \vec \nabla Y 
+
(\vec \nabla Y \wedge \vec \nabla X)\wedge \vec \nabla X
-
\frac{1}{T^2}
\vec E \cdot (\vec \nabla Y \wedge \vec \nabla X) (\vec E \wedge \vec \nabla X)
\right.\right.
\nonumber\\&& \left.\left.
+ \frac{1}{T^2}
(\vec B \cdot  \vec \nabla Y) \vec B +\frac{1}{T^2} \vec E \wedge (\vec E \wedge 
\vec \nabla Y)
\right)\right) = 0\nonumber\\
&&\vec \nabla \cdot \left( \frac{1}{R} \left( 
\phantom{\frac{1}{T^5}}\!\!\!\!\!\!\!\!\vec E 
+\vec \nabla X \wedge (\vec E \wedge 
\vec \nabla X)
+\vec \nabla 
Y \wedge (\vec E \wedge 
\vec \nabla Y)
\right. \right.\nonumber\\
&&
\left. \left.
+ \vec E \cdot (\vec \nabla Y \wedge  \vec \nabla X) (\vec \nabla Y \wedge
\vec \nabla X)
-\frac{1}{T^2} (\vec E \cdot \vec B) \vec B
\right)\right) =0
\nonumber\\
 &&
 \vec \nabla \wedge \left( \frac{1}{R} \left( \vec B +
  (\vec B \cdot 
\vec \nabla X) \vec \nabla X
+
(\vec B \cdot 
\vec \nabla Y) \vec \nabla Y
 -\frac{1}{T^2} (\vec E \cdot \vec B) \vec E
\right)\right)= 0 \nonumber\\
\label{tch2} 
\end{eqnarray}
where $R$ is now given by
\begin{eqnarray}
R^2 &=& 
           1
         +
           (\vec \nabla X)^2
         +         
           (\vec \nabla Y)^2
           + \left( \vec \nabla X\wedge \vec \nabla Y
           \right)^2
         +
           \frac{1}{T^2}
           \left(
           	    \vec B^2 
                  - 
                    \vec E^2
                  \right.  
           \nonumber\\
           && \left. 
           +
                    (\vec B \cdot \vec \nabla X)^2
                  -
                    (\vec E \wedge \vec \nabla X)^2
                    +(\vec B \cdot \vec \nabla Y)^2
                  -
                    (\vec E \wedge \vec \nabla Y)^2
            \right. \nonumber\\
           && \left. -(\vec E \cdot (\nabla X \wedge \vec \nabla Y))^2
           \right) - \frac{1}{T^4}
            (\vec E \cdot \vec B)^2  
\label{R2}
\end{eqnarray}

The solution takes the form
\begin{eqnarray}
\vec \nabla X   
 &= &({q_s^8}/{Tq_e}) \vec E \nonumber\\
 \, \vec \nabla Y  
 &= &({q_s^9}/{Tq_e}) \vec E 
\nonumber\\
\vec E &=& \frac{q_e}{4\pi\sqrt{r_0^4 + r^4}} \check r \nonumber\\
\vec B &=&  \frac{q_m}{4\pi r^2}  \check r
\label{dospikes}
\end{eqnarray}
where $q_s^8$ and $q_s^9$ are the charges of the two scalars, 
defined as in (\ref{lades}), 
\be
q_s^{a} \equiv T \int_{S_\infty}  dS_i \partial^i X^a  
\label{lades2}
\ee
and $r_0$ is now given by
\be
r_0^4 = q_e^2 + q_m^2 - \left(q_s^8\right)^2 - \left(q_s^8\right)^2
\label{asque}
\ee

This non-BPS solution can be interpreted as a spike that extends in
the direction $q_s^8 \check e^8 + q_s^9 \check e^9$,
with $\check e^8$ ($\check e^9$) denoting  the  unit vector in the
$X^8$ ($X^9$) directions. It is interesting to note that there is
a family of values for the scalar charges $({q_s^8}^{BPS},{q_s^9}^{BPS})$
which leads to the BPS limit,
\be
 \left({q_s^8}^{BPS}\right)^2 + \left({q_s^9}^{BPS}\right)^2 =  q_e^2 + q_m^2 
\label{camino}
 \ee
The particular solution ${q_s^8}^{BPS} = q_e$ and ${q_s^9}^{BPS} = q_m$
corresponds to the  BPS configuration with a $\nu = 1/4$ fraction
of unbroken supersymmetry  analysed in \cite{BLM}, which solves 
\be
\vec E = T \vec \nabla X   \, ,  \;\;\; \vec B = T \vec \nabla Y
\label{a}
\ee

It is interesting
to note that eqs.(\ref{a}) exhibit an invariance under transformations
which correspond to a  duality rotation for the electric and magnetic
fields and a related $SO(2)$ rotation for the scalar fields. Indeed, the
transformation 
\begin{eqnarray}
\vec E + i \vec B  &\longrightarrow & \exp(i \theta) 
(\vec E + i \vec B)\nonumber\\
X + i Y &\longrightarrow & \exp(i\theta) (X + i Y)
\label{sociologia}
\end{eqnarray}

Other solutions of eq.(\ref{camino}) are those that correpond
to $\nu = 1/2$, which    can be obtained by an $SO(2)$ rotation of
(\ref{grandiestadormido}) in the $(X^8,X^9)$ plane. 

\section{Dynamics and Boundary Conditions}
We shall now analyse the response of the theory to small fluctuation 
around
the static non-BPS dyon solutions that  we have found above, in the spirit of
ref.\cite{CM}.  

We take as a background the non-BPS spike solution (\ref{xab})    
and study the propagation of a $s$-wave  perturbation $\eta$, polarized
in a direction perpendicular
to the brane and to $\check X^9$, say $\check X^8$. Starting from action
(\ref{D3}), one obtains the linearized fluctuation equation
around the static solution
\begin{equation}
-\left(r^4 + \frac{q_e^2 + q_m^2}{(4\pi T)^2}
\right) \ddot \eta(r,t)+ 2r^3 \eta'(r,t) 
+ 
\left(
r^4 +  r_0^4
\right)\eta''(r,t) = 0
\label{sipero}
\ee
Writing $\eta(r,t) = \eta(r) \exp(i \omega t)$ 
and defining $x = \omega r$ the corresponding stationary equation is given by
\be
\frac{1}{x^2}f(x) (x^2 f(x) \eta'(x)
)'  + \frac{\kappa^2 + x^4}{x^4}\eta(x)  = 0
\label{lr}
\ee
where 
\begin{equation}
\kappa = \frac{\sqrt{q_e^2 + q_m^2}}{4\pi T}\omega^2
\end{equation}
 and
\be
f(x) = \frac{\sqrt{x^4 + \omega^4 r_0^4}}{x^2}
\label{seju}
\ee
In the BPS limit ($r_0 \to 0$) $f(x) \to 1$ and we recover the case discussed 
originally in \cite{CM},\cite{Svv}. 

To study eq.(\ref{lr}) we  change from $x$ to a new variable $\xi$
which
measures the length along $X$
\be
\xi(r)  =
\omega\int_{\sqrt{\kappa}/\omega}^r d\tilde r
\sqrt{1 +  {X'}^2(\tilde r)} 
\ee
Using the explicit form for the $X$ solution as given
in (\ref{xab}), $\xi$   can be written in the form
\be
\xi(x) = \int_{\sqrt \kappa}^x dy
\sqrt{\frac{y^4 + \kappa^2}{y^4 +  r_0^4 \omega^4}}
\label{zz}
\ee
Defining 
\be
\tilde \eta (x) = \left( x^4 +{\kappa^2}\right)^\frac{1}{4} \eta(x)
\label{zzz}
\ee
eq.(\ref{lr})   becomes a one-dimensional Schr\"{o}dinger  equation
\be
\left(- \frac{d^2}{d \xi^2}   + V(\xi)
\right) \tilde \eta(\xi) = \tilde \eta(\xi)
\label{cho}
\ee
with potential
\be
V(\xi) = \frac{5\kappa^2 x^6}{\left(x^4 + \kappa^2\right)^3}
+   \frac{(4 \pi T)^2 }{q_e^2 + q_m^2}r_0^4\kappa^2 x^2\frac{3\kappa^2 - 2 x^4}{\left(\kappa^2 + x^4
\right)^3}
\label{malv}
\ee
The first term in (\ref{malv}) is formally identical to the potential 
in the BPS limit \cite{CM}
except that the relation between $\xi$ and $x$, given by (\ref{zz}) depends
on $r_0$ and hence only coincides with the BPS answer for $r_0=0$. Another important
difference with the BPS case concerns the one dimensional domain in which
potential (\ref{malv}) is defined: being our solution a non-BPS 
one, $\xi$ extends from a finite (negative) $\xi(0)$ to $+\infty$, since
the cusp for this solution  has a finite height $X(0)$. Now, from 
$X^9= X(0)$ to infinity 
(i.e., in the $\xi$-interval $(-\infty,\xi(0))$) the 
disturbance just acts on the free scalar action of the semi-infinite string
attached to the brane. Then, in this region, one
has, instead of (\ref{cho}),
\be
- \frac{d^2 \tilde \eta(\xi)}{d \xi^2}    
  = \tilde \eta(\xi)
\label{chos}
\ee
We can then consider eq.(\ref{cho}) in the whole one dimensional domain by
defining 
\be
V_{eff}(\xi) = \left \{
\begin{array}{ll}
0 & \mbox{if $-\infty <\xi<\xi(0)$}
\\
V(\xi) & \mbox{if $ \xi(0) <\xi <\infty$}
\end{array}
\right.
\label{compv}
\ee

Potential (\ref{compv}), corresponding to a  non-BPS 
disturbed configuration, is more involved than the
BPS one, which was originally studied in the $\kappa \to 0$ limit 
using  delta function  and  square barrier approximations
\cite{CM},\cite{Svv}.  
We shall take this second way and approximate
 the potential by a rectangular potential,  adjusting its
height and width so that the integral of $V$ and the integral of $\sqrt V$
coincide. We then define
\be
S = \int \sqrt{V(\xi)} d\xi \, , \;\;\;\;  U = \sqrt \kappa\int V(\xi) d\xi
\label{ss}
\ee
One can see, by an appropriate change of variables that 
neither $S$ nor $U$ depend on $\kappa$. In terms of these quantities,
one finds  for the reflection  and transmission
amplitudes  
\begin{eqnarray}
R & = &  \frac{\exp \left(- i\sqrt \kappa S^2/U
\right)}{-1 + ({2i \sqrt \kappa S}/{U}) {\rm coth} S} \nonumber\\
T & = &  i\left(\frac{2\sqrt \kappa S}{U} {\rm cosech} S \right) R
\label{mex}
\end{eqnarray}
Eq. (\ref{mex}) shows that 
one has complete reflection  with a phase-shift approaching $\pi$
in the low-energy limit ($\kappa \to 0$). Computing numerically $S$ and $U$
one can also see that the non-BPS reflection coefficient $|R(r_0)|$ is  slightly
larger than the BPS one, $|R(r_0)| > |R(0)|$.

We thus conclude from the analysis above that a transverse disturbance on 
the string attached to the non-BPS brane,  reflects in agreement with the
expected result for Dirichlet boundary conditions: the reflection amplitude $R$
goes to $-1$ in the low-energy ($\kappa \to 0$) limit.  In the opposite limit
($\kappa \to \infty$) the potential vanishes so that the system passes from
perfectly reflecting to perfectly transparent at a scale
that, for the dyon background that we studied corresponds 
to  $ 8\kappa \sim 5 + r_0^4(4\pi T)^2/(q_e^2 + q_m^2)$. The emerging picture is then in agreement
with the D3-brane acting as a boundary for open strings. 

\section{Summary and discussion}
In summary we have constructed dyonic non-BPS solutions to the 
Dirac-Born-Infeld action for a U(1) gauge field in the
world volume coupled to one or two scalars
and analysed them in the context of brane dynamics. Although our solutions
also include those BPS ones already discussed in the 
literature, we have concentrated on the non-BPS sector to test whether 
this characteristic affects the picture of strings attached to branes. 
One important quantity in the analysis of the non-BPS solutions is the value
of the scalar charge $q_s$ which can be written in terms
of the electric and magnetic charge as 
\be
q_s^2 = q_e^2 + q_m^2 - (4\pi T)^2 r_0^4
\label{ul}
\ee
For $r_0^4 > 0$ our solutions
correspond to a brane with a spike
while for $r_0^4 < 0$ one has a brane-antibrane solution with a throat.
The  subtracted (renormalized) energy of these
non-BPS dyon solutions can be arranged in a way that
naturally leads to this picture of a brane pulled 
by a string with a tension $T_{(n,m)} = T\sqrt{n^2 + m^2/g_s^2}$ ($m$ and $n$
being the number of magnetic and electric flux units of the solution). As
shown graphically in Fig.4, as the scalar charge increases towards its
BPS value $q_s^{BPS}$, the spike grows and then, once $q_s^{BPS}$ is exceeded, 
the solution becomes a pair of brane-antibrane joined by a throat. Solutions
with two scalars can be constructed following analogous 
steps and also be interpreted in terms of spikes extending in the combined
direction of the two scalars.  

Finally we have studied the effect of small disturbances transverse both
the string and the non-BPS brane showing through a scattering analysis
that the results corresponds to the expected Dirichlet boundary conditions.
In particular, the reflection amplitude for the non-BPS background
is
slighty larger than the result for the BPS case and tends to $-1$ 
in the low-energy limit.

\vspace{2 cm}

\section*{Acknowledgements} This work is 
partially  supported by CICBA, CONICET (PIP 4330/96), ANPCYT
(PICT 97/2285), Argentina.  N.~Grandi is partially supported by
a  CICBA fellowship. G.~ Silva is supported by a CONICET fellowship.
H.R. Christiansen acknowledges partial support from FAPERJ, Funda\c c~ao de
Amparo \`a Pesquisa do Rio de Janeiro, Brazil.

\newpage

\begin{figure}
\centerline{ \psfig{figure=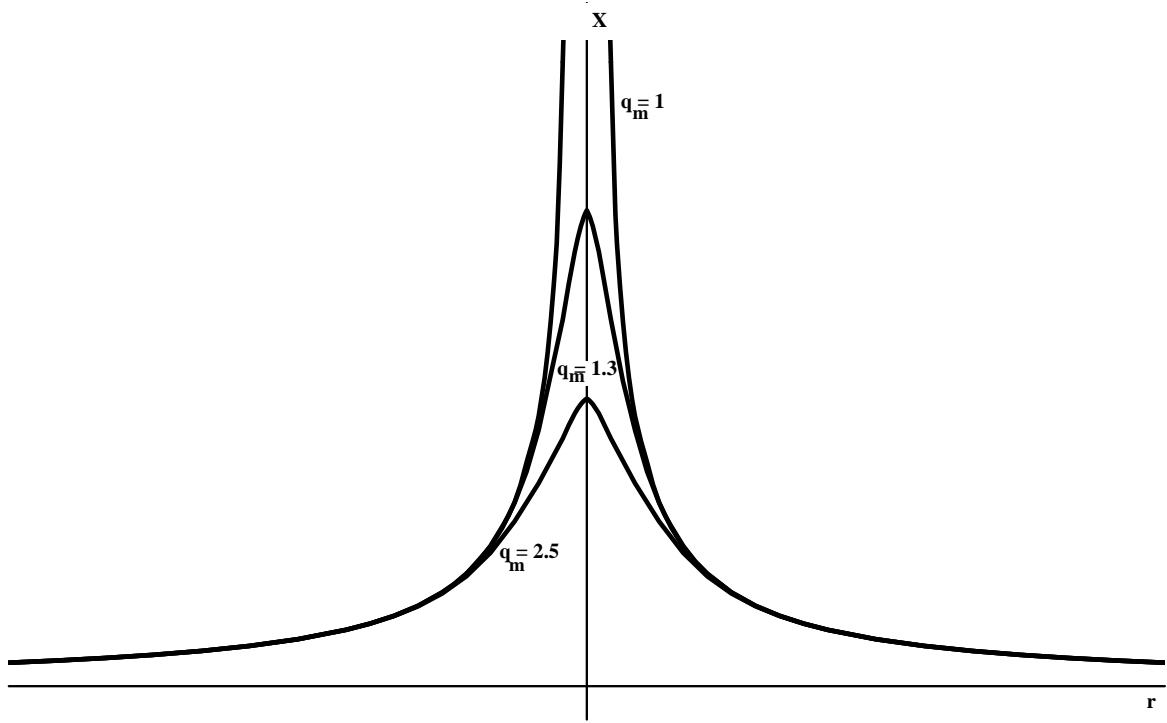,height=20cm,angle=0}}
\smallskip
\caption{ The scalar $X$  as a function of
$r$ for fixed electric charge $q_e=1$ and 
different values of the magnetic charge $q_m$. The value $q_m^{BPS}= 1$ 
is  the one for which the BPS limit is attained 
 when $q_e=1$ and $q_s= \sqrt 2$.
\label{fig-1} }
\end{figure}

\newpage

\begin{figure}
\centerline{ \psfig{figure=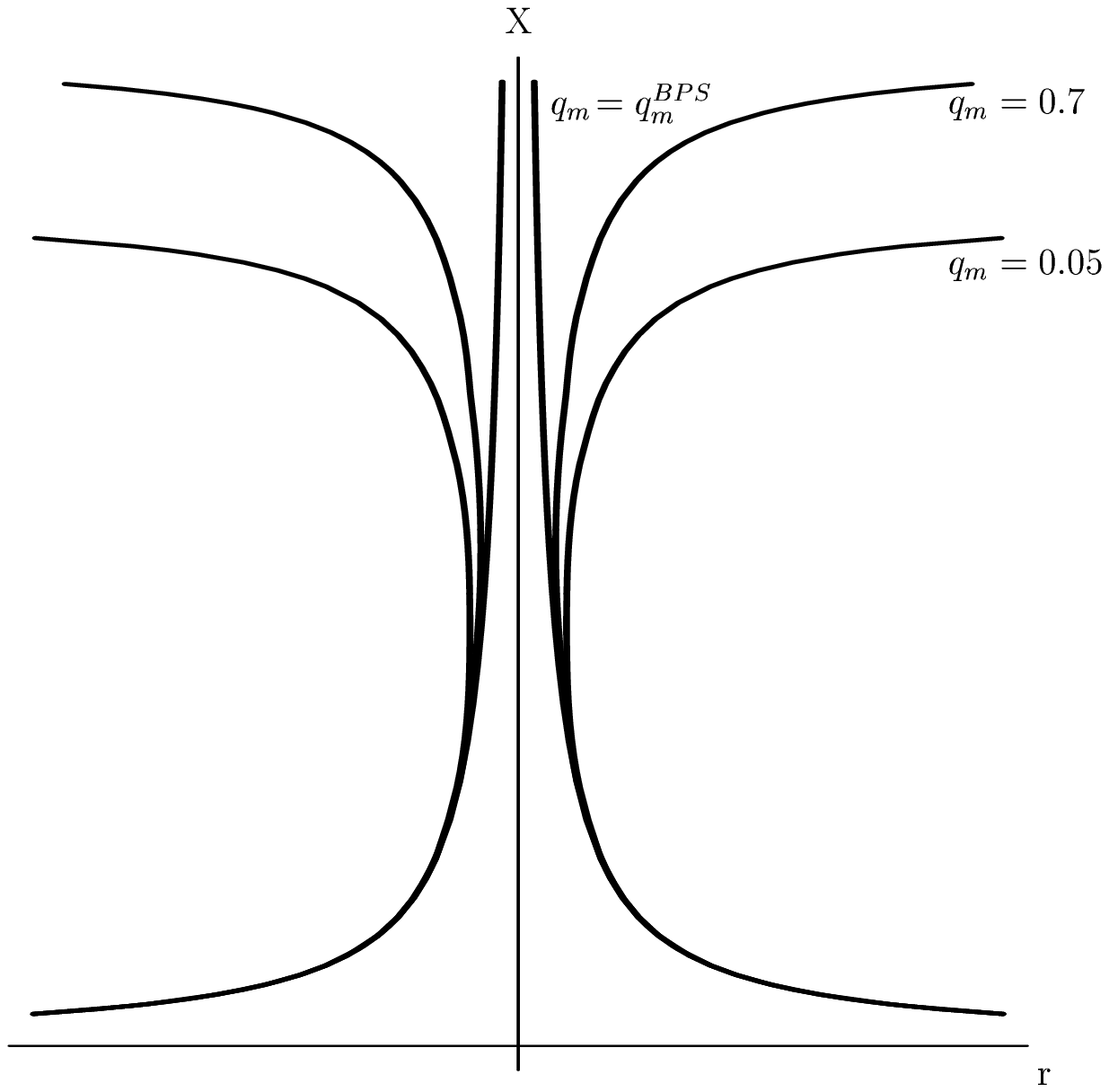,height=20cm,angle=0}}
\smallskip
\caption{The scalar $X$  as a function of
$r$ for fixed electric charge $q_e=1$ and 
different values of the magnetic charge $q_m$. The value $q_m^{BPS}= 1$ 
is  the one for which the BPS limit is attained 
 when $q_e=1$ and $q_s= \sqrt 2$. \label{fig-2} }
\end{figure}

\newpage

\begin{figure}
\centerline{ \psfig{figure=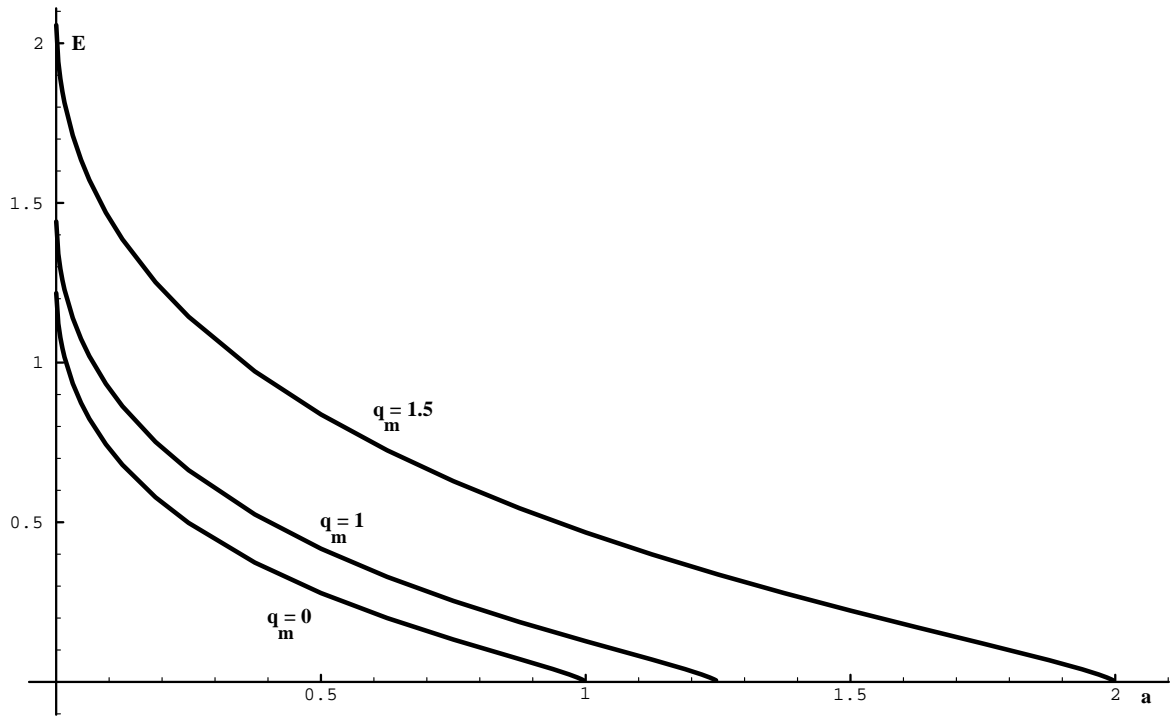,height=20cm,angle=0}}
\smallskip
\caption{The energy of the non-BPS configuration (a bent brane with
an attached string)  as a function of
$a = q_s^2/q_e^2$ for different values of the magnetic charge $q_m$.
  \label{fig-3} }
\end{figure}

\begin{figure}
\centerline{ \psfig{figure=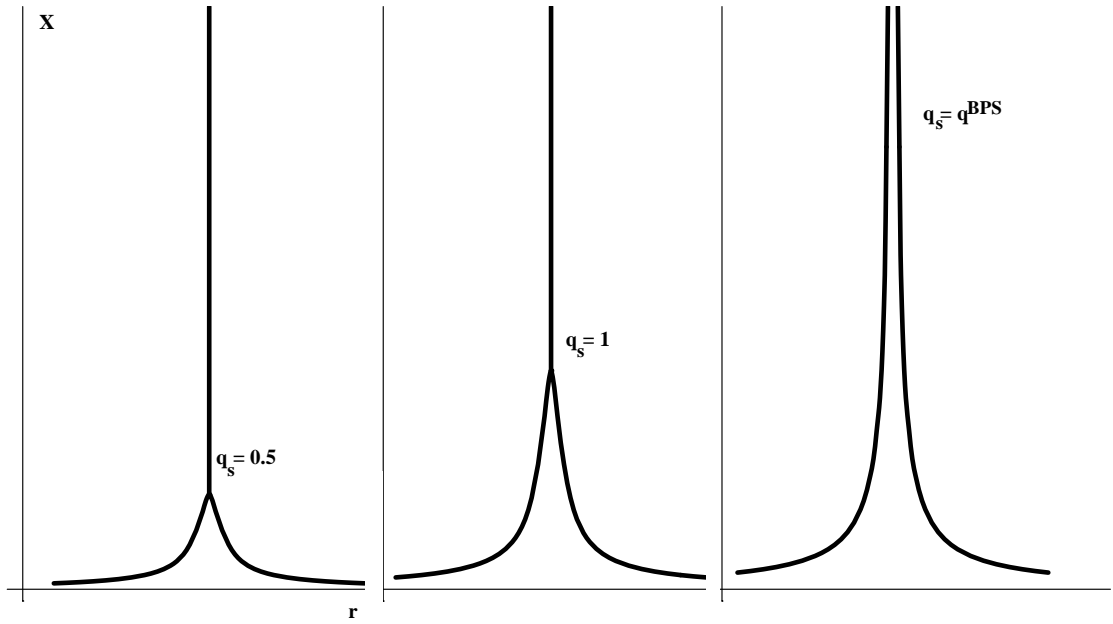,height=20cm,angle=0}}
\smallskip
\caption{ The spike in the brane, pulled by the string,
 as a function of $r$ for  values of
the scalar charge (in appropriate units)  going from $q_s = 0.5$ (left) 
to the BPS value $q_s^{BPS}= \sqrt{q_e^2 + q_m^2} = \sqrt 2$ (right).\label{fig-4} }
\end{figure}



\begin{references}
\bibitem{Pol} J.~Polchinski, Phys. Rev. Lett. {\bf 75} (1995) 4724.
\bibitem{Pol2} See J.~Polchinski, {\it Tasi Lectures on D-Branes},
hep-th/9611050 and references therein.
\bibitem{CM} C.~Callan and J.M.~Maldacena, Nucl.Phys. {\bf B513} (1998) 198
\bibitem{G} G.W.~Gibbons, Nucl. Phys. {\bf B514} (1998) 603.
\bibitem{GGT} J.P.~Gauntlett, J.~Gomis and P.K.~Townsend, 
JHEP {\bf 01} (1998) 003.
\bibitem{Hashi} A.~Hashimoto, Phys.Rev. {\bf D57} (1998) 6441. 
\bibitem{B} D.~Brecher, Phys. Lett. {\bf B442} (1998) 117.
\bibitem{Svv} G.K.~Savvidy, hep-th/9810163.  
\bibitem{Svv2} C.G.~Savvidy and K.G. Savvidy,  hep-th/9902023. 
\bibitem{BLM} D.~Bak, J.~Lee and H.~Min,
Phys.Rev. {\bf D59} (1999) 045011. 
\bibitem{KH} K.~Hashimoto, JEPTH {\bf 07} (1999) 016.
\bibitem{GKMTZ} J.P.~Gauntlett, C.~Koehl, D.~Mateos, P.K.~Townsend and 
M.~Zamblar, Phys. Rev. {\bf D60} (1999) 04504. 
\bibitem{g2} G.W.~Gibbons,   
Class.Quant.Grav. {\bf 16} (1999) 1471.
\bibitem{g3} G.W.~Gibbons  
{\it Lecture   at 6th Conference on 
Quantum Mechanics of Fundamental Systems, Chile,  1997}, 
hep-th/9801106.  
\bibitem{Ts} A.A. Tseytlin, to appear in the 
{\it Yuri Golfand memorial volume} ed. M. Shifman, Wd. Sci. , 2000, 
hep-th/9908105.
\bibitem{BG} O.~Bergman and M.R. Gaberdiel, hep-th/9908126.
\bibitem{Sch} J.H.~Schwarz, Phys.Lett. {\bf B360} (1995) 13.
\bibitem{Wi} E.~Witten, Nucl. Phys. {\bf B460} (1996) 335.
\bibitem{GK} K.~Ghoroku and K.~Kaneko, hep-th/9908154.
\bibitem{GR} G.~Gibbons and D.A.~Rasheed, Nucl. Phys. {\bf B454} (1995) 185.
%
%
\end{references}
\end{document}